\documentclass[prl,twocolumn,nofootinbib,aps,preprintnumbers,amsmath,amssymb,superscriptaddress,floatfix]{revtex4-1}
\usepackage{gensymb}
\usepackage{graphicx}
\usepackage{color}
\usepackage{bm}
\usepackage{amsmath}
\usepackage{xspace}
\usepackage[normalem]{ulem}
\usepackage{units}
\usepackage{dcolumn}
\usepackage{paralist}
\usepackage{natmove}
\usepackage{natbib}
\usepackage{epstopdf}%This line makes .eps figures into .pdf - please comment out if not required.

\definecolor{cream}{RGB}{222,217,201}
\newcommand{\HHCOOO}{H$_2$CO$_3$}
\newcommand{\COO}{CO$_2$}
\newcommand{\HCOOO}{HCO$_3^-$}
\newcommand{\COOO}{CO$_3^{2-}$}
\newcommand{\HOOO}{H$_3$O$^+$}
\newcommand{\OH}{OH$^-$}

\begin{document}

\makeatletter 
\newlength{\figrulesep} 
\setlength{\figrulesep}{0.5\textfloatsep} 

\newcommand{\topfigrule}{\vspace*{-1pt}% 
\noindent{\color{cream}\rule[-\figrulesep]{\columnwidth}{1.5pt}} }

\newcommand{\botfigrule}{\vspace*{-2pt}% 
\noindent{\color{cream}\rule[\figrulesep]{\columnwidth}{1.5pt}} }

\newcommand{\dblfigrule}{\vspace*{-1pt}% 
\noindent{\color{cream}\rule[-\figrulesep]{\textwidth}{1.5pt}} }

\makeatother

\title{Carbon Dioxide, Bicarbonate and Carbonate Ions in Aqueous Solutions at Deep Earth Conditions}
\author{Riccardo Dettori}
\affiliation{Department of Chemistry, University of California Davis, One Shields Ave. Davis, CA, 95616}
%\email{rdettori@ucdavis.edu}

\author{Davide Donadio}
\affiliation{Department of Chemistry, University of California Davis, One Shields Ave. Davis, CA, 95616}
\email{ddonadio@ucdavis.edu}

\date{\today}

\begin{abstract}
We investigate the effect of pressure, temperature and acidity on the composition of water-rich carbon-bearing fluids at thermodynamic conditions that correspond to the Earth's deep Crust and Upper Mantle. Our first-principles molecular dynamics simulations provide mechanistic insight into the hydration shell of carbon dioxide, bicarbonate and carbonate ions, and on the pathways of the acid/base reactions that convert these carbon species into one another in aqueous solutions. At temperature of 1000 K and higher  our simulations can sample the chemical equilibrium of these acid/base reactions, thus allowing us to estimate the chemical composition of diluted carbon dioxide and (bi)carbonate ions as a function of acidity and thermodynamic conditions. We find that, especially at the highest temperature, the acidity of the solution is essential to determine the stability domain of \COO\ vs \HCOOO\ vs \COOO. 
\end{abstract}
\maketitle

%%%MAIN TEXT%%%%
\section{Introduction}

Aqueous electrolyte solutions are an important component of geological fluids in the mid and deep Earth's crust and in the upper mantle,\cite{Thompson:1992ue} and water pockets may be present even in the transition zone at depths of more than 600 Km.\cite{Tschauner:2018ep}  
The chemical balance among different forms of ions in aqueous solution, and their pairing activity determine the formation and dissolution of minerals. At the extreme thermodynamic conditions of the Earth's deep crust and upper mantle water is supercritical and its different structural and physical properties, e.g. static dielectric constant and ionic conductivity, affect the relative stability and the structure of solvated species.\cite{Eckert:1996jc}

The Deep Carbon observatory (https://deepcarbon.net/) identified carbon dioxide dissolution and hydration in geological fluids as one of the Earth's most important reactions, as it affects the global carbon cycle.  Yet, the current molecular understanding of carbon dissolved in water-rich fluids at deep crust and upper mantle conditions is still limited in terms of both experiments and theoretical models. On the one hand, in experiments it is possible to probe solutions at extreme conditions by Raman spectroscopy, but the interpretation of such spectra is difficult and controversial.\cite{Hawke1974,Manning:2018jo} On the other hand, there are very few theoretical studies addressing carbonates in supercritical water by first principles:\cite{Pan:2016gb,Stolte:2019gr} the former addressing higher pressures and the latter focusing on the occurrence of carbonic acid as opposed in carbon dioxide-rich fluids. 

\begin{figure}[thb]
\centering
  \includegraphics[height=6cm]{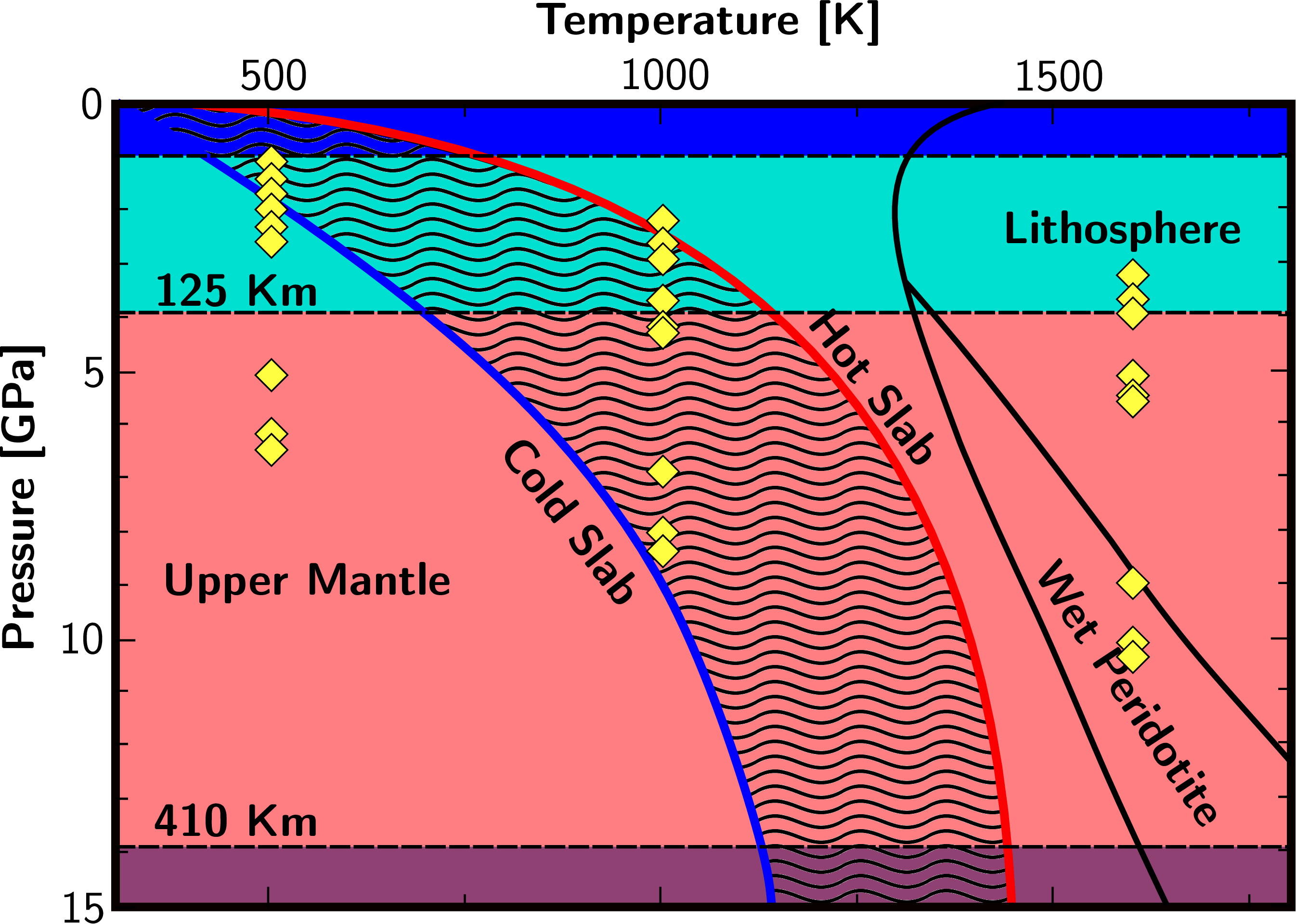}
  \caption{Continental (red solid line) and Oceanic (blue solid line) geotherms define the relevant range of pressure and temperature to study geological aquaeous solutions. Aqueous geofluids may be found also in the ``wet peridotite" domain. Yellow diamonds indicate the thermodynamic conditions of the calculations reported in this work. The graph is adapted from \citet*{Thompson:1992ue}. }
  \label{fgr:Fig1}
\end{figure}
The presence of aqueous electrolyte solutions along the Continental (hot) and Oceanic (cold) geotherms in the crust and upper mantle is significant at temperature up to 1400 K and pressure up to 14 GPa (Figure~\ref{fgr:Fig1}). Aqueous solutions may be also found in the "Wet Peridotite" region, at even higher temperatures, beyond 1600 K. 
Supercritical water behaves very differently as a solvent than at normal conditions. For example, the static dielectric constant of water varies from $\epsilon_{\rm w}\sim80$ at normal conditions to $\epsilon_{\rm w}\sim15$ at 1 GPa and 1000 K, and the autoionization constant of water ($\rm K_w$) increases by several orders of magnitude as a function of temperature and pressure.\cite{Holzapfel1969,Weingartner:2005jk,Pan:2013kb,Rozsa:2018kv}
The equations of state of minerals dissolved in supercritical water at geochemically relevant conditions are usually predicted using models fitted on  experimental data, such as the Helgeson-Kirkham-Flowers (HKF) model \cite{Helgeson:1981cm} or its recent development, the Deep Earth Water (DEW) model.\cite{Sverjensky:2014ie} Both HKF and DEW models provide an estimate of the free energy of solvation of ions ($\Delta G_s$) as a function of the dielectric constant of water $\epsilon_{\rm w}$ and of an electrostatic Born parameter specific to each ion, such that:
\begin{equation}
\Delta G = -\dfrac{N_Az^2e^2}{8\pi\epsilon_0r_0}\left(1-\dfrac{1}{\epsilon_{\rm w}}\right),
\label{eq:DEW}
\end{equation}
where $z$ is the ion charge, $e$ the electron charge, $r_0$ effective radius of the ion, $\epsilon_0$ the vacuum permittivity and $N_A$ is the Avogadro number. 
While the DEW model is considered predictive and has wide use in the geochemistry community, its parameters are fitted to experiments at mild conditions, and the validity of the extrapolations to extreme conditions lacks compelling verification. In fact, experimental data of $\epsilon_w$ for water at elevated temperatures are currently limited to 550$\degree$ C and 0.5 GPa.\cite{Sverjensky:2014ie, Caciagli2003, Martinez2004} 

First principles molecular dynamics (FPMD) simulations, based on density functional theory (DFT), proved valuable to make up for the experimental gap providing a parameter-free estimate of the dielectric constant of supercritical water for pressure up to 10 GPa.\cite{Pan:2013kb} This achievement suggests that a systematic use of FPMD will allow geochemists to extend aqueous geochemical models to a broader range of thermodynamic conditions. 
\textcolor{black}{In addition, recent refinements of the standard Born models of solvation\cite{Nakamura2018,Duan2015} call for a more accurate description of the solvent polarization in the presence of ions, which can be provided by FPMD simulations.}
Ultimately, molecular simulations can provide a direct prediction of the stability domain of aqueous electrolytes at geologically relevant thermodynamic conditions. 
Especially in the cases of anomalous behavior, atomistic insight from FPMD simulations would rationalize experimental observations, made possible by the development of techniques to probe liquids at extreme conditions, such as high-pressure Raman\cite{Hawke1974} and NMR.\cite{Pautler:2014ix}

FPMD describes accurately the properties of water and ice at high pressure \cite{Schwegler:2000jt,Schwegler:2001ja,Schwegler2008,Goldman:2009kb,Pan:2013kb,Rozsa:2018kv} and it has been employed in the past to study the solvation structure of various aqueous species \cite{Tuckerman2002}, from monoatomic anions and cations,\cite{Todorova2008, Gaiduk2017, Pham2016} to complex molecules of biological importance.\cite{Gaigeot2003} 
\textcolor{black}{FPMD is particularly useful to study reactions in solutions and at conditions at the limit of experimental possibilities, not only at high pressure and temperature but also in the presence of strong static electric field.\cite{Pietrucci:2015jq,Cassone2019}}
Specifically relevant to the present study, FPMD has already been used to resolve the solvation shell of carbon dioxide and the conversion of \COO \ into bicarbonate in high-pH environment,\cite{Leung2007, Zukowski2017} showing that this computational technique is able to quantitatively predict free energies of reaction as well as reaction barriers, provided that suitable correction terms are calculated from higher-level gas phase quantum chemical calculations.\cite{Grifoni:2019eu} 
%%\textcolor{black}{Besides , FPMD was successfully used to investigate the effect on water structure and dynamics under the action of extreme static electric field\cite{Cassone2019}, once again proving its reliability in probing conditions at the limit of experimental possibilities}.
Using parameter-free FPMD simulations a pioneering work by Pan and Galli\cite{Pan:2016gb} showed that, contrary to conventional models, at Upper Mantle conditions (1000 K and 11 GPa) \COO \ is not the major carbon species, but it transforms into \COOO \ and \HCOOO. 
Intermediate pressures, corresponding to the boundary between the Lithosphere and the Upper Mantle were studied to unravel the abundant occurrence of \HHCOOO. However, a systematic study of the carbon species in water-rich fluids at these conditions is still lacking, and the effects of the acidity of the solutions remain largely unexplored.  
In this work, we perform extensive FPMD simulations to investigate the properties of dissolved carbon in geological fluids at the deep crust and upper Mantle conditions as a function of temperature, pressure and acidity (Figure~\ref{fgr:Fig1}). To set different pH conditions we performed simulations of $\sim$1 molar solutions with different initial conditions, namely 2Na$^+ +$\COOO, Na$^++$\HCOOO \ and \COO. 
Low-temperature simulations shed light on the solvation shell of the three carbon species as a function of pressure.
At temperatures at and above 1000 K the carbon species are highly reactive and FPMD simulations allow us to map the average composition of the solutions at both  acidic and  basic conditions as a function of pressure and temperature.
Furthermore, we obtain direct insight into the hydration and dehydration reaction pathways of carbon solutes, without the bias of either {\it a priori} assumptions or the choice of specific reaction coordinates. 

%We show the structure of these systems by computing the radial distribution function and their dynamical behavior by computing the mole percents of different carbon species as functions of simulation time. Furthermore, we provide a first estimation of the acidity of carbon aqueous solutions at extreme conditions.

\begin{figure*}[htb]
\centering
  \includegraphics[width=16cm]{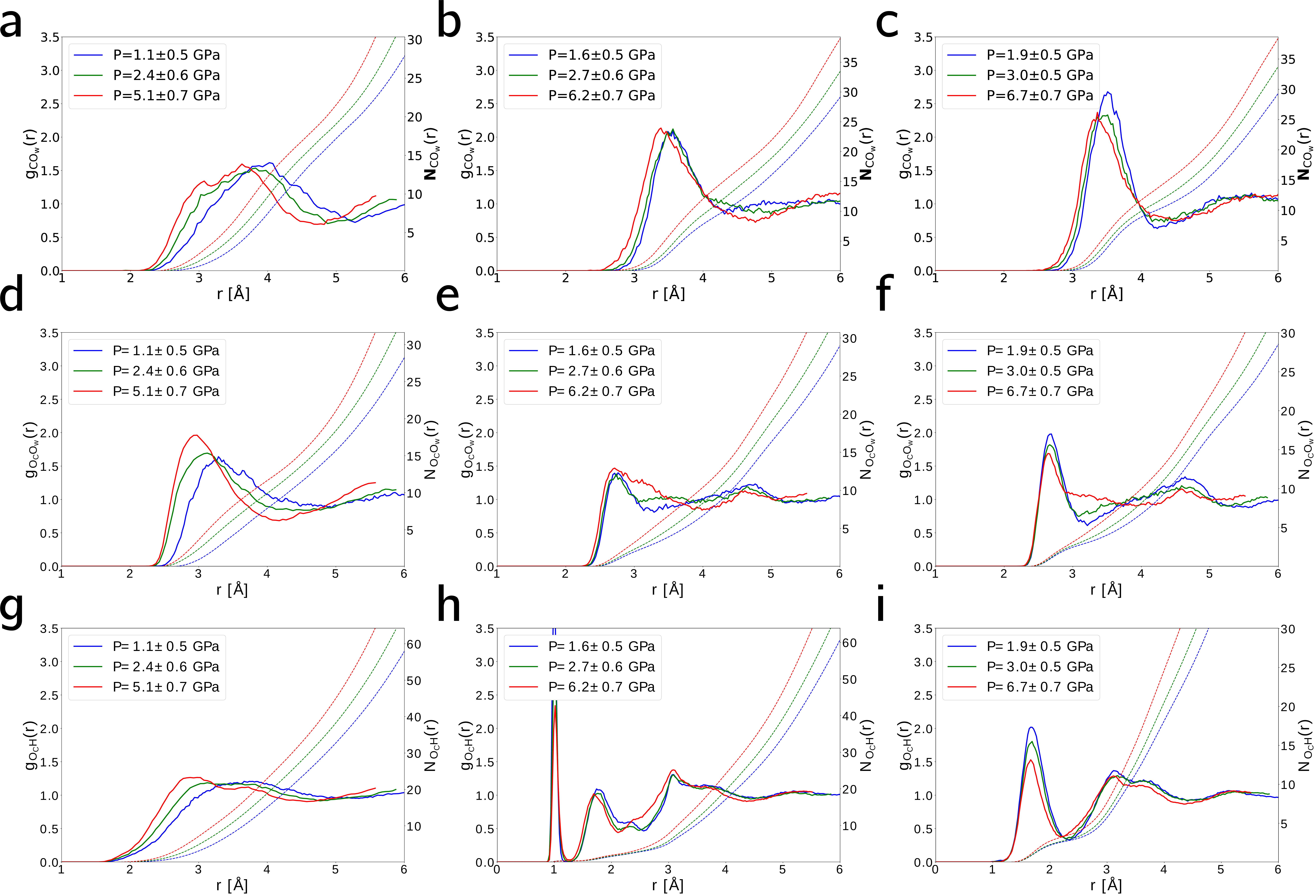}
  \caption{\ Solvation shells of \COO \ (a, d, g), \HCOOO \ (b, e, h), and \COOO \ (c,f, i) at $T=500$~K, characterized by the carbon-oxygen radial distribution function and integrated number of neighbors at different pressures (a, b, c), by the oxygen-oxygen radial distribution function (d, e, f) and by the oxygen-hydrogen radial distribution function (g, h, i). An atomistic representation of the solvation shell for each carbon species is shown in the Supplementary Information$^\dag$, in Fig. S1. Models of solvation shells at different temperatures and pressure are shown in Fig. S2}
  \label{fgr:Fig2}
\end{figure*}

\section{Methods and Models}

FPMD simulations were carried out using the Quickstep approach, implemented in the \texttt{CP2k} code.\cite{Vandevondele2005} 
We employ the Perdew-Burke-Ernzerhof (PBE) generalized gradient approximation (GGA) for the DFT exchange and correlation functional.\cite{Perdew1996} 
Valence Kohn-Sham orbitals are represented on a Gaussian triple-$\zeta$ basis set with polarization, optimized for molecular systems in gas and condensed phase,\cite{VandeVondele2007} and core states are treated implicitly using Geodecker-Teter-Hutter pseudopotentials.\cite{Goedecker1996} 
\textcolor{black}{Semilocal GGA functionals do not provide a good description of dispersion forces, and it is well established that including van der Waals corrections improves the description of water at ambient temperature.\cite{Lin:2012es,Jonchiere:2011fr} However, the effect of vdW corrections, and manybody effects in general, in supercritical water is less drastic and it has not been studied in detail for the PBE functional.\cite{Jonchiere:2011fr,Schienbein:2017fa}}
In fact, the PBE functional has been extensively used to study the properties of supercritical water \cite{Pan:2013kb, Rozsa:2018kv, Schwegler2008} and of hydrocarbons at deep Earth conditions as well.\cite{Spanu2011} 
Although it is known that GGA may not be accurate in the description of solvated doubly-charged anions because of the charge delocalization error,\cite{Cohen2008} it was recently shown that PBE performs better under extreme conditions than under ambient conditions in terms of equation of state of water, as well as in predicting static and electronic dielectric constants.\cite{Pan:2013kb, Rozsa:2018kv} Pan and Galli also showed that PBE provides very similar results for the structure and the speciation of \COO\ and \COOO\ in supercritical water.\cite{Pan:2016gb,Stolte:2019gr}

When carbon species are dissolved in water, the following hydration/dehydration reactions and their equilibrium constants determine the relative composition of the solution:
\begin{align}
\label{eqn:reactions}
\begin{split}
 \rm{CO_2 + 2H_2O \leftrightharpoons HCO_3^- + H_3O^+} \\
 \rm{HCO_3^- + H_2O \leftrightharpoons CO_3^{2-} + H_3O^+} \\
 \rm{HCO_3^-(aq) \leftrightharpoons CO_2 + OH^-} \\ 
 \rm{CO_3^{2-} + H_2O \leftrightharpoons HCO_3^- + OH^-}
\end{split}
\end{align}
To probe these reactions for diluted solutions, we prepared cubic simulation cells containing 63 water molecules and one solvated species (\COO, \HCOOO, or \COOO ) with edge length $L=12.37$ \AA\ with periodic boundary conditions. Charge neutrality is achieved by compensating the net negative charge of the anions by a corresponding number of sodium cations. This initial conditions correspond to total densities between 1.03 and 1.09 g/cm$^{3}$. 
To perform simulations at higher pressure, we reduced the box linear size to $11.76$ and $11.14$ \AA. These systems correspond to solution with different molar concentrations: specifically 0.88 $M$, 1.02 $M$, and 1.20 $M$ respectively. 
For each system, we performed a set of MD simulations at $T=500$, $1000$, and $1600$ K. The MD equations of motion are integrated with the velocity Verlet integrator with a timestep of 0.25 fs. To perform simulations in the constant temperature constant volume canonical ensemble (NVT) we used the stochastic velocity rescaling thermostat,\cite{Bussi2007} with a coupling constant $\tau=0.5$ ps. 
Each model was first equilibrated for 5 ps at the target temperature, and the analysis was performed on data obtained from 50 ps-long production runs. The overall project amounted to 27 runs for a total simulation time of $\sim1.5$~ns. 

If the reactions in Eq.~\ref{eqn:reactions} can be considered at equilibrium over time scales of several tens of picoseconds, we can exploit direct FPMD simulations to estimate the chemical balance of the solvated species at given thermodynamic conditions.
Since the products of these reactions are hydronium and hydroxide ions, from sufficiently long simulations we can also obtain the equilibrium concentrations of these species, which allow us to roughly estimate the pH and pOH of the solution. In principle, to fix the pH, one would have to perform grand canonical (GC) ensemble simulations with a fixed chemical potential for either H$_3$O$^+$ or OH$^-$. However, the standard approach to GC simulations, i.e. Widom particle insertion, is extremely costly and practically unfeasible for condensed phases, especially with first-principles simulations.\cite{Widom:1963fl,FrenkelSmit} Yet, here we argue that it can be avoided by fixing the conjugate thermodynamic variable of the chemical potential, i.e. the number of particles in the system. This idea works in the same way as fixing the volume (or the density) of a system determines the equilibrium pressure, and fixing the energy determines the equilibrium temperature. 
Therefore, in reactive simulations we can achieve the goal of mapping the relative abundance of the solute species as a function of temperature, pressure and acidity. 
Using the concentrations of \HOOO\ and \OH\ averaged over a production run, and the experimental value of the autoionization constant of supercritical water,\cite{Holzapfel1969,Hamann:1969fk} we estimate the acidity of the solution as in the textbook case of weak polyprotic acids and bases. We define the acidity/basicity of the solution as the difference between pH and pOH (see Supporting Information). 
However, given the small size of the simulation cell, the high solute concentration and the limited accessible timescale, this estimate is hampered by large uncertainties. More accurate estimates of acidity and solutes stability may be achieved either simulating larger systems, or computing directly redox potentials by free energy calculations, for example exploiting the Born-Haber cycle,\cite{Cheng:2014jb} which is, however, beyond the scope of this work.

\section{Results and Discussion}

\subsection{Solvation of Carbon Species at 500 K}

We carried out a first set of low-temperature simulations  to characterize the solvation shell of the different carbon species as a function of pressure at $T=500$ K. As expected, these systems show no reactivity on the  timescale of our production runs, thus no information about the predominant species can be extracted from these simulations. Geochemical models and previous experiments suggest that at relatively low temperature and pressure, the major dissolved carbon species is \COO(aq)\cite{Manning:2018jo,Duan2006,Zhang2009,HUIZENGA2001}. Nevertheless, the metastability of all three carbon solutes allows us to shed light into the structure of their hydration shell, which has an essential role in the formation of \textcolor{black}{minerals, such as calcite or dolomite,} at mild conditions,\cite{Raiteri:2010kd,DeLaPierre:2017jp} and for carbon geosequestration.\cite{Liang:2017ih}
Further studies using enhanced sampling methods would be required to estimate the relative stability of dissolved carbon species and the composition of solutions in the colder layers of the Earth's crust.\cite{Grifoni:2019eu} 

Figure~\ref{fgr:Fig2}(a,d,g) shows that the first solvation shell of \COO\ is rather unstructured, giving rise to a broad first peak in the carbon-oxygen radial distribution  function (RDF) ($g_{\rm CO}(r)$), as well as in the oxygen-hydrogen RDF ($g_{\rm O_CH}(r)$, where $O_c$ indicates the oxygen atom bonded to carbon). Also the oxygen-oxygen RDF exhibits a broad first peaks and weak structuring at low pressure, while the peak gets better defined and shifts to shorter distance at higher pressure (Fig.~\ref{fgr:Fig2}d).  
The reason is that \COO\ is apolar and does not form hydrogen bonds with the surrounding water molecules. This behavior, observed in both experiments and simulations at ambient conditions,\cite{Gallet:2012vh,Zukowski2017} is retained at 500 K for pressures ranging from 1.1 to 5.1 GPa. The effect of increasing pressure is to shift the first broad peak of $g_{\rm CO}(r)$ toward smaller distances, indicating a spatial contraction of the first solvation shell, but no significant change in the number of nearest neighbors, which is defined as the integral of the first peak of the $g_{\rm CO}(r)$.     

The carbon-oxygen RDFs of anionic species, \HCOOO, and \COOO, (Figure~\ref{fgr:Fig2}b,c) exhibit a sharper first peak that shifts toward smaller distance as a function of pressure. Both anions form hydrogen bonds with water, as it appears from the well-defined structure of the $g_{\rm O_CH}(r)$ in Fig.~\ref{fgr:Fig2}(h,i).
The OH group of \HCOOO\ donates one H-bond, and each oxygen accepts on average 2.2 H-bonds from the neighboring water molecules.\footnote{\textcolor{black}{The number of hydrogen bonds has been calculated according to the same geometric criteria used in Ref.\cite{Luzar}: two molecules are considered hydrogen bonded if the oxygen-oxygen distance is lower than 3.3 \AA, the oxygen-(donor)hydrogen distance is lower than 2.4 \AA and the H-O$\ldots$O angle is smaller than 30$^o$.}} This number is not sensitive to pressure, but the overall hydrogen-bonding structure of the solvation shell undergoes significant changes when the pressure is increased from 2.7 to 6.2 GPa. 
At high pressure a larger number of water molecules from the second solvation shell enters the radius of the first shell of hydrogen-bonded molecules, disrupting the local order of the hydrogen-bonded network. This effect can be observed also for \COOO: the double-charged anion accepts on average 2.8 H-bonds per oxygen, leading to a total number of $\sim$9 neighbors, in agreement with former FPMD simulations,\cite{Leung2007} but smaller than that estimated at ambient conditions with empirical potentials.\cite{Bruneval:2007ik,Gale:2011ila,Raiteri:2015bq}
Increasing pressure causes weakening of hydrogen bonding, as showcased by the lower peak of the $g_{\rm OH}(r)$ (Figure~\ref{fgr:Fig2}h and i), but an overall increase of the neighboring water molecules results in a more disordered and dynamic solvation shell. 

%Such solvation shell is consistent with previous FPMD simulation at ambient conditions\cite{Leung2007,Gallet:2012vh,Zukowski2017}. As for the bicarbonate and carbonate ions, the solvation shell is more packed than \COO \ and it becomes slightly more dense as the pressure increase, as shown by the number of the nearest neighbors. Such effect is due to the polar nature of the carbon complex which promotes the formations of an increasing number of hydrogen bonds.

\subsection{Solvation of Carbon Species at 1000 and 1600~K}

High temperature and high pressure enhance the rate of autoionization in water, thus making the solvent increasingly reactive. At $T=1000$ K and pressures above 11 GPa, water rapidly dissociates and recombines through a bimolecular mechanism that produces short-lived hydronium-hydroxide ion pairs, and it eventually turns into an ionic fluid.\cite{Goncharov:2005gi,Goldman:2009kb,Rozsa:2018kv} 
Conversely, at lower pressure along the Hugoniot compression curve, nearly no autoionization was observed in pure water over the typical FPMD simulation timescale of few tens ps.\cite{Schwegler:2000jt,Schwegler:2001ja}  
In our simulations at $T=1000$ K we also do not see spontaneous water autoionization up to  $P\sim 4$ GPa, whereas at higher pressure transient H$_3$O$^+$/OH$^-$ pairs occur spontaneously.
Nevertheless, in all the runs, except for those starting with \COO\ at $P=2$ and 3.9~GPa, the enhanced reactivity of supercritical water engenders fast interconversion among carbon dioxide, bicarbonate and carbonate ions with the consequent release of either hydronium or hydroxide ions, according to the reactions in Eq.~\ref{eqn:reactions}. 

\begin{figure}[th]
\begin{center}
\includegraphics[width=0.78\columnwidth]{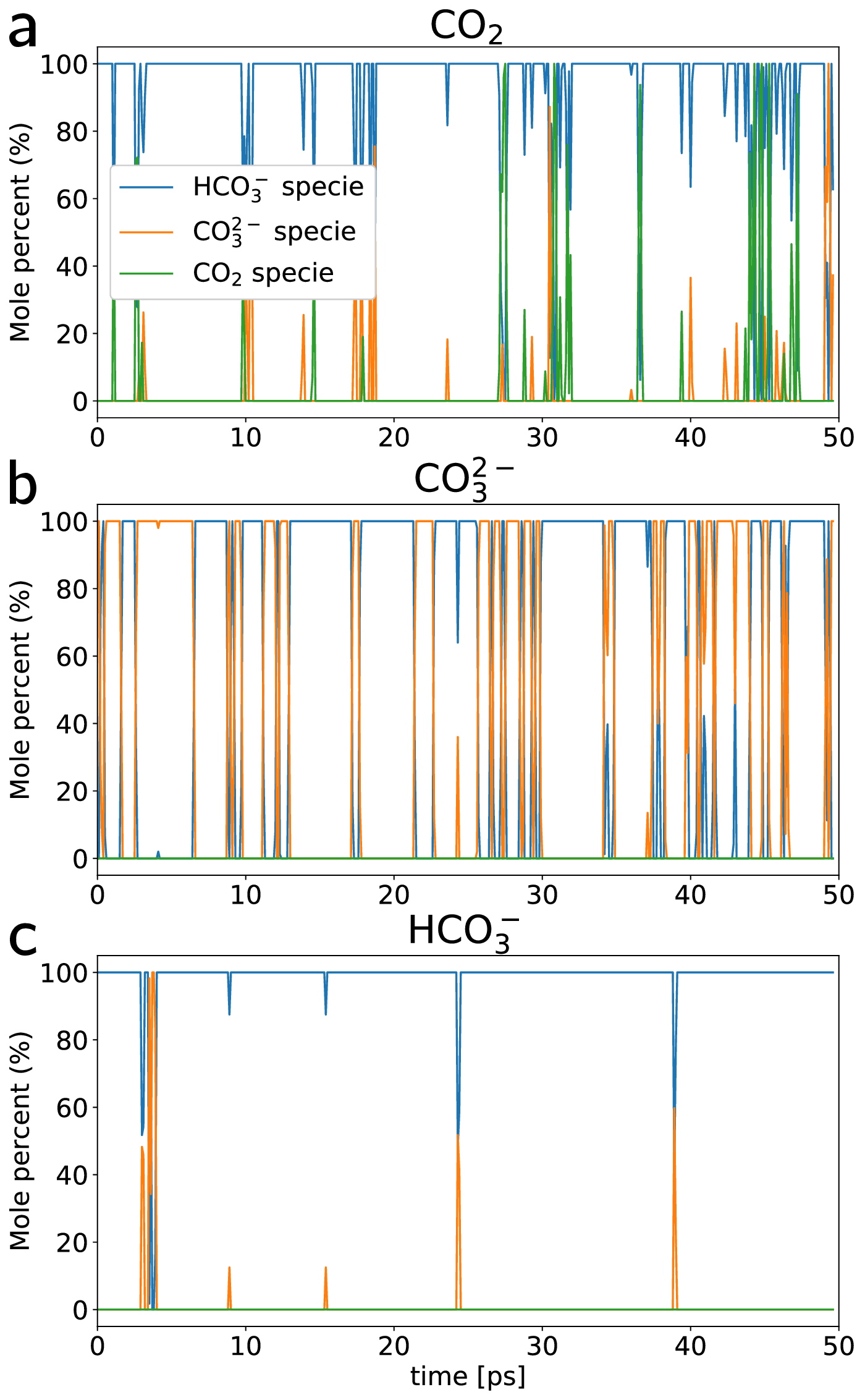}
\caption{Mole percents of \COO \ (green), \COOO \ (orange), and \HCOOO \ (blue) calculated as functions of time at $T=1000$K with \COO \ (a), \COOO \ (b), and \HCOOO  \ (c) as the starting solvated species. Here the mole percent relative to \COO \ actually considers also the presence of \HHCOOO \ because carbonic acid appears in  amounts comparable to \COO. The reaction $\rm CO_2 + H_2O \leftrightharpoons H_2CO_3$ does not alter the pH of the solution.}
\label{fgr:species}
\end{center}
\end{figure}

Figure~\ref{fgr:species} shows an example of the evolution of the carbon solutes in the three runs, with \COO, \COOO\  and \HCOOO\ as the starting solvated species, at 1000 K and at the smallest cell volume, which leads to pressures between 6.9 and 8.3 GPa. In this analysis the three species are identified using a geometric criterion involving the C-O distance and the number of oxygen atoms bonded to carbon. The cutoff distance to count an oxygen-carbon bond is set to 1.5 \AA\ and the number of bonds allows us to single out \COO\ from the two ionic species. 
If three oxygen atoms are bonded to the carbon atom, we use the $\rm O-H$ distance ($< 1.2$~\AA) to determine whether the solute molecule is either \COOO \ or \HCOOO.
In the same way as Ref.\cite{Pan:2016gb}, we define the percent molar fraction of a given species $i$ at time $t$ as $ x_i(t)=\dfrac{n_i(t)}{N}\times100\% $, where $n_i(t)$ is the number of steps containing the $i$th species between the time $(t-\tau)$ and $t$, and $N$ is the total number of snapshots in this time interval. We set the time interval $\tau$ to 50 fs.

%%%%%% REACTION PATHS

\begin{figure}[htp]
\begin{center}
\includegraphics[width=0.95\columnwidth]{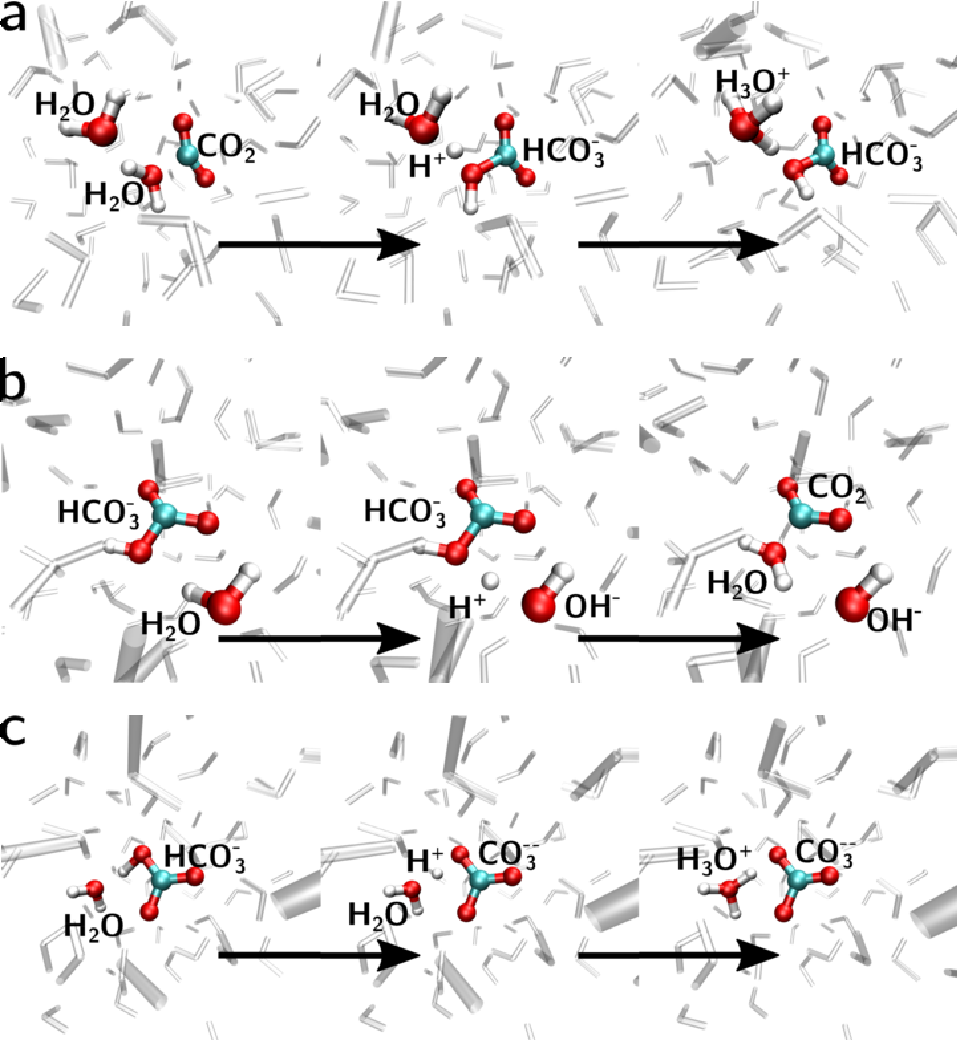}
\caption{Atomistic representation of the mechanism of the reactions that transforms \COO\ into \HCOOO\ (a), \HCOOO\ into \COO\ (b)  and \HCOOO\ into \COOO\ (c). The first and the third reactions are mediated by water in the first solvation shell and release one hydronium ion in solution, while the second one produce one hydroxide ion.}
\label{fgr:reactions}
\end{center}
\end{figure} 
The spontaneous hydration/dehydration reactions among \COO, \HCOOO and \COOO\ in unbiased FPMD simulations allows us to explore the reaction mechanisms at the atomic scale. Figure~\ref{fgr:reactions} shows the representative molecular pathways of the water-assisted transformation of \COO\ into \HCOOO, \HCOOO\ into \COOO, and of \HCOOO\ into \COOO. 
The formation of bicarbonate from carbon dioxide neither require the presence of a free hydroxide ion\cite{Leung2007} nor has \HHCOOO\ as an intermediate. The reaction occurs via the nucleophilic attack of a water molecule to \COO. The oxygen atom of the H$_2$O molecule attacks the carbon of \COO, while one of its protons is released into the solution, initially forming a \HOOO \ ion with a neighboring water (Fig.~\ref{fgr:reactions}a). Eventually the proton diffuses in solution via Grotthuss mechanism.  
Also the reverse reaction, from \HCOOO\ to \COO, proceeds along a  water-mediated pathway, and we never observe direct breaking of the C-OH bond. The OH group of \HCOOO\ accepts a proton from a neighboring water molecule and detaches from the carbon atom as a water molecule (Figure~\ref{fgr:reactions}b). The H$_2$O that donated the proton diffuses in solution as a hydroxide ion.
Both the conversion of  \HCOOO\ into \COOO\ (Figure~\ref{fgr:reactions}c) and the reverse reaction (not shown) occur via direct proton exchange with a neighboring water molecule. 
It is important to stress that all these reactions do not require the direct interaction between the carbon species and either \OH\ or \HOOO. \textcolor{black}{Identifying these mechanistic reaction pathways is an essential step to eventually implement enhanced sampling methods to refine the calculation of reaction free energies.\cite{Pietrucci:2015jq,Grifoni:2019eu}}

%%%%%
\begin{figure}[bt]
\centering
  \includegraphics[width=0.95\columnwidth]{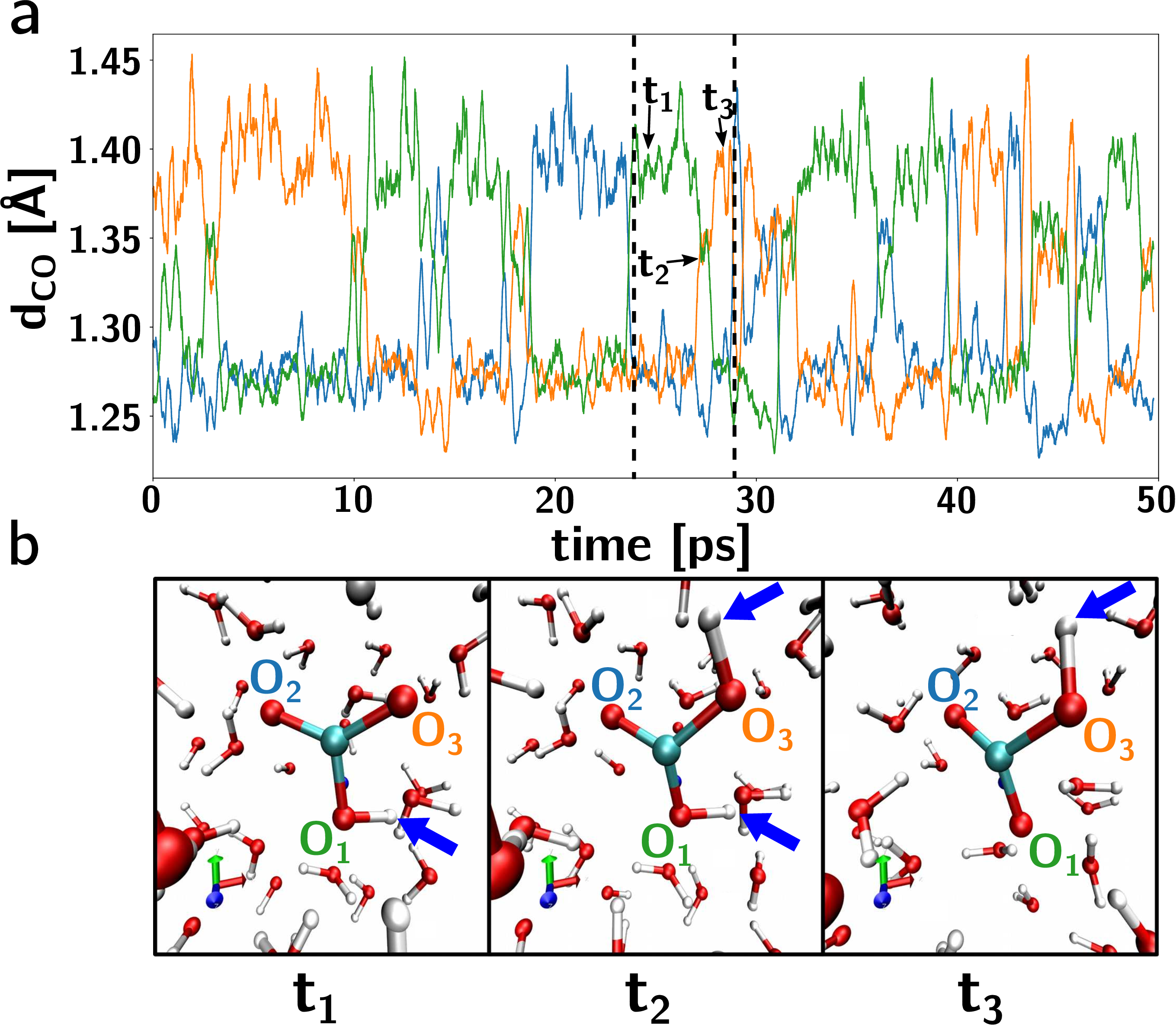}
  \caption{
  (a) Carbon-oxygen distances calculated over the productions runs starting with solvated \COO\  at P=6.9 GPa and T=1000 K. The predominant species at these conditions is \HCOOO\ but the position of the OH group frequently changes from one oxygen site to another through a reaction mediated by the nearest neighboring molecules, with the occurrence of \HHCOOO (b). \textcolor{black}{In the presence of \HCOOO \ the average $r_{\rm CO}$ for the protonated oxygen atom is larger ($\sim 1.4$ \AA) than the other two C-O bonds ($\sim 1.28$ \AA), while in the case of \HHCOOO, the average $r_{\rm CO}$ for the two protonated oxygen atoms is slightly reduced ($\sim 1.35$ \AA). Accordingly, we used black arrows to highlight the three different steps for such swapping mechanism. Blue arrows indicate the hydrogen atoms involved in the process.}}
  \label{fgr:BondSwap}
\end{figure}

Besides the reactions given in Eq.~\ref{eqn:reactions}, we observe that the configuration of \HCOOO\ is highly dynamic, in the sense that the OH site swaps from one oxygen to another. This mechanism can be tracked by monitoring the length of the C-O bond length $(r_{\rm CO})$ for each oxygen atom in  \HCOOO.  The average $r_{\rm CO}$ for the protonated oxygen atom is larger ($\sim 1.4$ \AA) than the other two C-O bonds ($\sim 1.28$ \AA). Figure~\ref{fgr:BondSwap}a shows that over a trajectory of a solution mostly containing \HCOOO\  the "long" C-O bond evolves from one site to another through rapid transitions. 
Figure~\ref{fgr:BondSwap}b shows how this swapping mechanism happens: a water molecule hydrogen bonded to an oxygen of the bicarbonate donates a proton to the ion forming transient neutral carbonic acid (central panel in Fig.~\ref{fgr:BondSwap}b). 
The abundance of \HHCOOO\ at these conditions is actually relevant and comparable to that of \COO\ in agreement with extensive simulations recently performed by Stolte and Pan.\cite{Stolte:2019gr} 
In the following analysis we group the concentrations of \COO\ and \HHCOOO\ into a single neutral contribution, as the relative abundance of one of these two neutral species with respect to the other does not affect the acidity of the solution.

At higher temperatures ($T=1600$ K) the solutions exhibit further increased reactivity than those at 1000 K, with frequent transitions from one species to another, thus providing more accurate statistic on the equilibrium composition of the fluid. In these high temperature runs the solvent remains mostly molecular, and the main transition mechanisms are the same as those described in detail for the simulations, involving neutral water molecules as facilitators of the reactions. However, in the high pressure runs ($P=10$ GPa) we observed a more significant ionic character of supercritical water, possibly enhanced by the presence of bicarbonate and carbonate ions.  

Comparing the carbon-oxygen radial distribution functions at different pressure and temperature (Figure S2 in the Supporting Information), we observe that the effect of the pressure at high temperature is similar to that discussed for the runs at 500 K, and the structure of the first solvation shell of \COO, \HCOOO\ and \COOO\ does not change significantly with temperature.

%%%%%%%%%%%% ACIDITY MAPS explanation and discussion %%%%%%%%%%
\subsection{Composition of Carbon-Bearing Fluids as a Function of Pressure and Acidity}

The relatively high frequency of reactive events allows us to consider the chemical reactions at equilibrium and to estimate the equilibrium molar fraction of solutions at given thermodynamic conditions as the average molar fraction of over a whole production run. It is important to note that, as one can infer from Figure~\ref{fgr:species}, the average compositions of the solution at similar temperature and pressure may differ significantly, depending on the starting solute species. The reason is that the hydronium/hydroxide ions produced in the acid/base reaction among different solutes change the pH of the solutions. The latter can be estimated in the same way as the molar fraction, by averaging the concentration of excess hydronium/hydroxide over a trajectory. 
Simulations starting with \COO\ result in acidic solutions, as the transformation into \HCOOO \ produces hydronium, unless no reaction happens and \COO\ remains the only solvated species for the whole duration of the run.
According to the same argument, systems starting with \COOO\ can only probe basic conditions, whereas \HCOOO\ may transform producing either \HOOO \ or \OH, thus engendering either acidic or basic equilibrium conditions.

The autoionization constant of supercritical water $\rm K_w$ is extremely sensitive to temperature and pressure and it increases monotonically with either. In the $(P,T)$ range explored in this work it ranges from $\rm K_w = 10^{-7}$ at $P=1$ GPa and $T=1000$ K to $\rm K_w \gtrsim 10^{-1}$ at $P=10$ GPa and $T=1600$ K.\cite{Holzapfel1969} Hence, we cannot use pH alone to define acidity or basicity of a supercritical solution, as the condition of neutrality changes with the change of $\rm K_w$. We then classify the acidity/basicity of the solutions in terms of the difference between pOH and pH, which corresponds to the $\log_{10}$ of ratio between the equilibrium concentrations of \HOOO\ and \OH ($f=\rm{pH}-\rm{pOH}=\rm -log_{10}[H_3O^+]/[OH^-] $), computed from the reactive FPMD simulations.
%, reporting $\rm{pH-pOH} = \rm -log_{10}[H_3O^+]/[OH^-]$. 
$\rm{pH}-\rm{pOH}=0$ defines neutrality at any temperature and pressure, $f=\rm{pH}-\rm{pOH}<0$ represents acid conditions and $f=\rm{pH}-\rm{pOH}>0$ basic conditions. A detailed explanation about the calculation of such ratio is reported in Supplementary Information$^\dag$.

\begin{figure}[htp]
\begin{center}
\includegraphics[width=0.98\columnwidth]{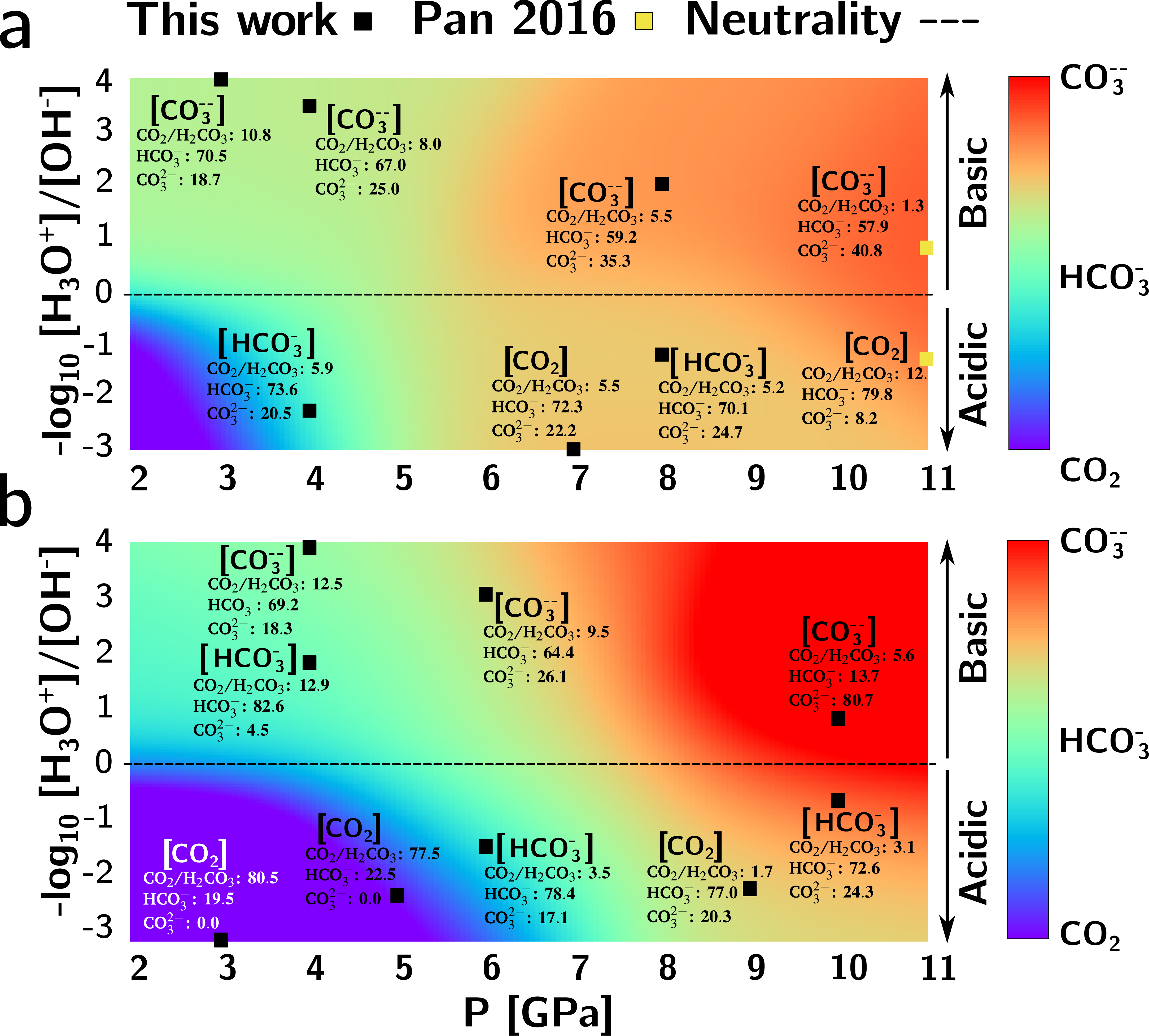}
\caption{Composition of carbon species in aqueous solution as a function of pH and pressure, for $T=1000$K (a) and $T=1600$K (b). In square brackets is reported the initial carbon species. The top panel contains also the first results at extreme conditions obtained by FPMD simulations.\cite{Pan:2016gb} The mole percent, the pressure and the $f$ values can be found in the Supplementary Information$^\dag$ in Table S1 and S2.}
\label{fgr:maps}
\end{center}
\end{figure} 
In Figure~\ref{fgr:maps} The composition of the water-rich carbon-bearing fluid is mapped as a function of pressure and the acidity parameter defined above at T=1000 K (a) and 1600 K (b).\footnote{For reference, in water at ambient conditions \COO\ is stable at acidic conditions, for pH$<$6.5, \HCOOO\ for pH from 6.5 to 10.5 and \COOO\ at basic conditions pH$>$10.5.}  
For each point the graph reports in brackets the initial species and, below, the equilibrium composition averaged over the 50 ps FPMD run. In the 1000 K plot we include also two high-pressure points from Ref.~\cite{Pan:2016gb} which adopts a similar computational framework as ours and yields results in good agreement. Moreover, we omit the low and intermediate pressure simulations starting from \COO, as they do not exhibit any reactivity, and we are not able to verify whether such \COO\ stability is the consequence of still too high free energy barrier for the \COO--\HCOOO reaction. 
At 1000 K the solution is dominated by the presence of \HCOOO\ which is the most abundant species at all the conditions probed. At acidic conditions there is still a significant amount of \COO\ up to 7 GPa (14\%), but nearly none remains at higher pressures in the range of acidity and overall solute concentration considered here. At basic conditions \COOO\ is relatively present, and its abundance increases from about 19\% to above 40\% in mole percent as a function of pressure. 

At higher temperature ($T=1600$K, Figure~\ref{fgr:maps}b) the range of composition is wider. At acidic conditions there is an broad region, where \COO/\HHCOOO \ are the most abundant species, that extends up to $\sim 5$ GPa. At higher pressure \HCOOO\ takes over as the dominant species up to 10 GPa. 
At neutral and basic conditions \HCOOO\ is the most abundant species up to $\sim 7$ GPa. At higher pressure (10 GPa) the basic fluid contains almost entirely \COOO\ (90\%).

%\textcolor{magenta}{Here we discuss the maps and the trends!}

%\textcolor{magenta}{(Revised this part)
%Higher temperatures ($T=1600$K) shift the aforementioned reactions towards the production of ions at even lower pressures, causing zero or very rare appearance of aqueous carbon dioxide.  }

The observed trends in the relative stability of the three carbon species may be qualitatively interpreted in terms of the DEW model and Eq.~\ref{eq:DEW}. Former FPMD simulations show that the static dielectric constant of supercritical water increases with pressure and decreases as a function of temperature.\cite{Pan:2013kb} 
Our calculations suggest a similar trend in the stability of \COO/\HHCOOO: pressure destabilizes carbon dioxide in favor of bicarbonate and eventually of carbonate. Conversely, higher temperature stabilizes \COO/\HHCOOO \ at low and intermediate pressure.
Surprisingly, high temperatures shift to the right the equilibrium of the reaction that produces \COOO\ from \HCOOO\ at high pressure, reducing the window of stability of the bicarbonate ion, especially in a basic environment. This latter effect may stem from the transition of water from molecular to mostly ionic, as the water autoionization constant at these conditions approaches the unit value.\cite{Holzapfel1969} 
Supporting this interpretation, Figure S3 shows that the mean square displacements of water oxygen and hydrogen at low pressure/low temperature overlap within statistical uncertainty, as it happens for molecular fluids.
Conversely, at $P=10$ GPa and $T=1600$ K the two curves have different slopes, indicating that there is a substantial amount of free ions in solution.

\section{Conclusions}

In conclusion, we carried out an extensive set of FPMD simulations of a water-rich carbon-bearing fluid at various temperatures and pressures corresponding to the thermodynamic conditions of the Earth's deep crust and upper mantle ($2<P<10$ GPa and $500<T<1600$ K).
We systematically probed the effect of preparing the systems with carbon species as solutes, namely \COO, \HCOOO, and \COOO \ at $\sim 1$ M molar concentration, and we found that in highly reactive conditions, averaging over sufficiently long runs  gives an unbiased estimate not only of the composition of the solution, but also of its acidity. 
This observation allows us to map the composition of geological fluids as a function of temperature, pressure and acidity: this information is essential for geochemical models, as the stability of different forms of carbon solutes impacts the growth and dissolution of \textcolor{black}{calcium, magnesium and iron carbonates, including calcite, aragonite, dolomite, magnesite and siderite, which are present in significant amounts both in the crust and in the mantle.\cite{Berg:1986gi,Manning:2018jo}} 
Our simulations provide direct evidence that, as opposed to what is customarily assumed in geochemical models, \COO(aq) is not the major carbon species present in water-rich geological fluids in the Earth's deep crust and upper mantle.\cite{Pan:2016gb,Duan2006,Zhang2009,HUIZENGA2001,Stolte:2019gr}
Nevertheless, we find that the equilibrium composition of the solutions depends critically on the initial conditions, which, in turn determine the equilibrium content of \HOOO \ and \OH \ ions. This result highlights the importance of considering acidity, at the same level as temperature and pressure, to predict the composition of geological fluids at deep Earth conditions.

The HKF and DEW models can account qualitatively for the observed trends in the relative stability of different carbon species as a function of pressure and temperature. However, at high temperature (1600 K) and pressure ($\sim 10$ GPa), as water approaches the transition from molecular to ionic, the increasing abundance of free \HOOO \ and \OH \ ions influences the reactivity of the system and favors the formation of \COOO, at mildly basic conditions.

Furthermore, FPMD simulations also provide insight into the atomistic mechanisms of the protonation/deprotonation reactions of different carbon species. These reaction pathways are non-trivial and mostly involve neutral water molecules in the first solvation shell of the carbon species, and proton-hopping. For example, we find that \COO\ transforms into \HCOOO\ via the nucleophilic attack of a neutral water molecule and the release of hydronium. Conversely the reverse reaction proceeds via the electrophilic attack of the OH group and the concurrent release of a water molecule and a hydroxide ion.

%We considered different carbon species in solutions at temperature/pressure conditions of the Earth's deep crust and upper mantle and we investigated the effect of the proton-exchange reactions that control the molecular composition affecting the acidity of the solution. 

%Our results not only confirm previous findings, but expand present data availability providing a substantial amount of information at thermodynamics conditions, which are very difficult to probe experimentally. We found that especially at high pressures \COO(aq) is the least abundant carbon species in supercritical water, and that by increasing the temperature the equilibrium between \HCOOO \ and \COOO \ becomes very sensitive to pressure. Considering the data we obtained, we were able to provide a first estimation of aqueous solutions containing carbon under several different conditions, mapping such composition as a function of temperature, pressure and acidity.

\section*{Conflicts of interest}
The authors declare that they have no competing interests.

\section*{Acknowledgements}
We are grateful to Prof. William H. Casey for useful discussions and critical reading of the manuscript.
%%%END OF MAIN TEXT%%%


%%%REFERENCES%%%
\bibliography{rsc} %You need to replace "rsc" on this line with the name of your .bib file
\bibliographystyle{rsc} %the RSC's .bst file

\end{document}